\documentclass[10pt,conference]{IEEEtran}

\usepackage{siunitx}
\usepackage{bm}
\usepackage{graphicx} 
\usepackage{float}
\usepackage{cite}
\usepackage{amsmath,amssymb}
\usepackage[outdir=./]{epstopdf}
\usepackage{afterpage}
\usepackage{optidef}
\usepackage{orcidlink}
\usepackage{soul}
\usepackage{tikz}
\usepackage{fontawesome5} 
\usetikzlibrary{calc}
\usepackage{pgfplots}
\usetikzlibrary{shapes.geometric, fit, positioning, backgrounds, arrows.meta}
\usepackage{multirow}
\usepackage{subcaption}

\IEEEoverridecommandlockouts 

\usepackage{hyperref}  
\hypersetup{
	colorlinks=false,
	linkbordercolor=white,
	urlbordercolor=white,
	pdfborder={0 0 0}
}  
\usepackage{adjustbox}

\makeatletter
\ams@newcommand{\iiiiint}{\DOTSI\protect\MultiIntegral{5}}
\renewcommand{\MultiIntegral}[1]{%
  \edef\ints@c{\noexpand\intop
    \ifnum#1=\z@\noexpand\intdots@\else\noexpand\intkern@\fi
    \ifnum#1>\tw@\noexpand\intop\noexpand\intkern@\fi
    \ifnum#1>\thr@@\noexpand\intop\noexpand\intkern@\fi
    \ifnum#1>4 \noexpand\intop\noexpand\intkern@\fi 
    \noexpand\intop
    \noexpand\ilimits@
  }%
  \futurelet\@let@token\ints@a
}
\makeatother

\begin{document}
\bstctlcite{IEEEexample:BSTcontrol}

\title{Cooperative and Collaborative Multi-Task Semantic Communication for Distributed Sources \thanks{This work was supported in part by the German Ministry of Education and Research (BMBF) under Grant 16KISK016 (Open6GGub) and the German Research Foundation (DFG) under grant 500260669 (SCIL).}}

\author{
\IEEEauthorblockN{Ahmad Halimi Razlighi\,\orcidlink{0009-0006-3826-832X}, Maximilian H. V. Tillmann\,\orcidlink{0009-0000-6548-4278}, \\ Edgar Beck\,\orcidlink{0000-0003-2213-9727}, Carsten Bockelmann\,\orcidlink{0000-0002-8501-7324}, and Armin Dekorsy\,\orcidlink{0000-0002-5790-1470}}\\
\IEEEauthorblockA{Department of Communications Engineering, University of Bremen, Germany}
\IEEEauthorblockA{E-mails:\{halimi, tillmann, beck, bockelmann, dekorsy\}@ant.uni-bremen.de}
}

\maketitle

\begin{abstract}

In this paper, we explore a multi-task semantic communication (SemCom) system for distributed sources, extending the existing focus on collaborative single-task execution. We build on the cooperative multi-task processing introduced in \cite{10654356}, which divides the encoder into a common unit (CU) and multiple specific units (SUs). While earlier studies in multi-task SemCom focused on full observation settings, our research explores a more realistic case where only distributed partial observations are available, such as in a production line monitored by multiple sensing nodes. To address this, we propose an SemCom system that supports multi-task processing through cooperation on the transmitter side via split structure and collaboration on the receiver side. We have used an information-theoretic perspective with variational approximations for our end-to-end data-driven approach. Simulation results demonstrate that the proposed cooperative and collaborative multi-task (CCMT) SemCom system significantly improves task execution accuracy, particularly in complex datasets, if the noise introduced from the communication channel is not limiting the task performance too much. Our findings contribute to a SemCom framework capable of handling distributed sources and multiple tasks simultaneously, advancing the applicability of SemCom systems in real-world scenarios.

\end{abstract}


\section{Introduction} \label{section.Intro}
Recent breakthroughs in artificial intelligence, particularly in deep learning (DL) and end-to-end (E2E) communication technologies, have led to the rise of semantic communication (SemCom) \cite{Gunduz2022}. It has attracted significant attention, being recognized as a critical enabler for the sixth generation (6G) of wireless communication networks. SemCom is expected to play a key role in supporting a wide range of innovative applications that will define 6G connectivity and beyond \cite{WenTong}.

In contrast to conventional communication systems, which are designed based on Shannon's information theory and focus on the accurate transmission of symbols, SemCom prioritizes understanding the meaning and goals behind transmitted information. SemCom operates at the second level of communication, the semantic level, where the goal is to convey the desired meaning rather than ensuring exact bit-level accuracy \cite{Sana2022}. By surpassing the traditional focus on the precise transmission of bits, SemCom is well-suited for emerging applications, such as the industrial internet and autonomous systems, where successful task execution is prioritized over the exact reconstruction of transmitted data at the receiver.

Research into SemCom has explored five main approaches, with four detailed in \cite{Wheeler2023} and a fifth inspired by Weaver's extension of Shannon’s theory to include the semantic level \cite{weaver1953recent}. These approaches are:
\begin{itemize}
    \item  Classical approach: Quantifies semantic information using logical probability.
    \item Knowledge graph (KG) approach: Represents semantics through structured KGs.
    \item Machine learning (ML) approach: Encodes semantics within learned model parameters.
    \item Significance approach: Focuses on timing as a key component of semantic meaning.
    \item Information theory approach: Extends Shannon's framework to address semantic-level communication.
\end{itemize} 

Recent works in SemCom primarily focus on two research directions: data reconstruction and task execution. ‌‌‌‌Initial investigations into data recovery were led by \cite{Xie2021} and \cite{Xie2021-2}, which utilized ML techniques to reconstruct diverse data sources such as text, speech, and images. Building on these foundational works, \cite{Yan2022} and \cite{Tong2021} have extended the focus to explore concepts like communication efficiency in SemCom. In addition, systems dealing with structured data have been examined through the KG approach to enhance data recovery \cite{9685056KG}.
\begin{figure}[!t]
    \centering
    \resizebox{0.5\textwidth}{!}{\input{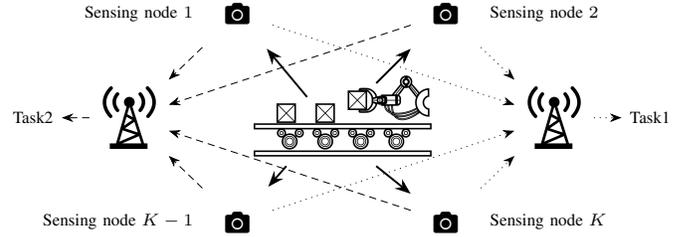}}
    \caption{An example of cooperative and collaborative multi-tasking for distributed sources.}
    \label{fig:intro_system_model}
\end{figure}

In task-oriented SemCom, the focus shifts to executing intelligent tasks at the receivers. Most research in this area has concentrated on single-task scenarios. For example, \cite{Shao2021} developed a communication scheme using the information bottleneck framework, which encodes information while adapting to dynamic channel conditions. 

Moreover, some works considered more realistic scenarios in which the source is distributed, for instance, \cite{Shao2022} studied distributed relevant information encoding for collaborative feature extraction to fulfill a task.\cite{Beck2023} also offered a framework for collaborative retrieval of the message using multiple received semantic information.

To address practical communication scenarios, SemCom systems must be capable of handling multiple tasks simultaneously. Early efforts, such as \cite{xie2022task} and \cite{he2022learning}, explored non-cooperative methods where each task operates on its respective dataset independently. Conversely, recent works like \cite{10013075}, \cite{10520522}, and \cite{gong2023scalable} studied joint multi-tasking using established ML approaches and architecture.

The prior works on multi-tasking in SemCom have focused on ML approaches, however, \cite{10654356} introduced an information-theoretic analysis of the problem. This study proposed a split structure for the semantic encoder, dividing the semantic encoder into a common unit (CU) and multiple specific units (SUs), to enable cooperative processing of various tasks. The split structure can perform multiple tasks based on a single observation. In this work, we have examined a more applicable scenario, in which the full observation is not accessible, however, we have different distributed views of our main observation. This is illustrated in Fig. \ref{fig:intro_system_model}, where the full observation is the whole view of the production line but our sensing nodes provide partial observations from different positions.

In distributed cases, where the source/observation is distributed, the task cannot be executed depending on a single sensing node alone. Thus, collaborative systems, where nodes jointly complete a single task at the receiver side, have been studied in \cite{Beck2023} and \cite{Shao2022}. However, our research expands the multi-task scenarios to distributed settings, by combining the split structure from \cite{10654356}, which brings cooperation of multiple tasks on the transmitter side, with collaboration on the receiver side. Therefore, we have contributed to a SemCom system capable of cooperatively and collaboratively executing multiple tasks by proposing a cooperative and collaborative multi-task (CCMT) SemCom architecture. In our data-driven approach, we have tailored semantic communication to multi-task processing for distributed sources and formulated it through information theory using variational approximations. Key contributions include:

\begin{itemize}
    \item Combining the cooperative multi-task process, enabled by the split structure, with the collaborative process of the distributed observations.
    \item Considering different training methods and channel conditions for the proposed CCMT architecture.
    \item Demonstrating the effectiveness of the CCMT system by showing enhancements in task execution over various scenarios, specifically when dealing with more challenging datasets.
\end{itemize}
\section{System Model} \label{section.SystemModel}
This section introduces our probabilistic modeling of the proposed CCMT model. Furthermore, we formulate an information-theoretic optimization problem that aims to optimize the joint execution of multiple tasks in an E2E manner.
\subsection{System Probabilistic Modeling}\label{subsec.system_prob_model}
\begin{figure}[!t]
    \centering
    \resizebox{0.45\textwidth}{!}{\begin{tikzpicture}[>=stealth, node distance=1.5cm]
        \node[text=red] (G) {};
        \node[right=3 of G, text=red] (Z) {};
        \node[right=0.32 of G] (B) {};
        \node[left=0.3 of Z] (Q) {};
        \node[right=2.5 of Z, yshift=0.3 cm] (S) {$\begin{tabular}{@{}c@{}}
                                                    \text{Observation} \\
                                                    \small \text{($\bm{S}$)}
                                                  \end{tabular}$};
        \draw[fill=white] (G.center)++(0.8,0.8) circle (0.5) node[font=\scriptsize] (GCn) {$\text{Task}_N$};
        \draw[fill=white] (Z.center)++(0.8,0.8) circle (0.5) node[font=\small] (ZCn) {$z_N$};
        \node[right=0.13 of G, xshift=0.8 cm, yshift=0.8 cm] (Bn) {};
        \node[left=0.1 of Z, xshift=0.8 cm, yshift=0.8 cm] (Qn) {};
        \draw[->] (Bn) -- (Qn);
        \draw[fill=white] (G.center)++(0.3,0.3) circle (0.6) node (GC3) {};
        \draw[fill=white] (Z.center)++(0.3,0.3) circle (0.6) node (ZC3) {};
        \node[right=0.23 of G, xshift=0.3 cm, yshift=0.3 cm] (B2) {};
        \node[left=0.3 of Z, xshift=0.3 cm, yshift=0.3 cm] (Q2) {};
        \draw[->] (B2) -- (Q2);
        \draw[fill=white] (G.center)++(0.15,0.15) circle (0.65) node (GC2) {};
        \draw[fill=white] (Z.center)++(0.15,0.15) circle (0.65) node (ZC2) {};
        \node[right=0.28 of G, xshift=0.15 cm, yshift=0.15 cm] (B1) {};
        \node[left=0.3 of Z, xshift=0.15 cm, yshift=0.15 cm] (Q1) {};
        \draw[->] (B1) -- (Q1); 
        \draw[fill=white] (G.center) circle (0.7) node[font=\small] (GC1) {$\text{Task 1}$};
        \draw[fill=white] (Z.center) circle (0.7) node[font=\large] (ZC1) {$z_1$};
        \draw[->] (B) -- (Q);

        \fill (G.center)++(2.0,0.5) circle [radius=0.7pt];
        \fill (G.center)++(2.07,0.57) circle [radius=0.7pt];
        \fill (G.center)++(2.14,0.64) circle [radius=0.7pt];

        \fill (G.center)++(5.05,0.5) circle [radius=0.7pt];
        \fill (G.center)++(5.12,0.57) circle [radius=0.7pt];
        \fill (G.center)++(5.19,0.64) circle [radius=0.7pt];

        \node[right=0.32 of Z] (QQ) {};
        \node[right=2.47 cm of Z] (BB) {};
        \draw[->] (QQ) -- (BB);
        \node[right=0.29 of Z, xshift=0.15 cm, yshift=0.15 cm] (QQ1) {};
        \node[right=2.33 of Z, xshift=0.15 cm, yshift=0.15 cm] (BB1) {};
        \draw[->] (QQ1) -- (BB1);
        \node[right=0.24 of Z, xshift=0.3 cm, yshift=0.3 cm] (QQ2) {};
        \node[right=2.18 of Z, xshift=0.3 cm, yshift=0.3 cm] (BB2) {};
        \draw[->] (QQ2) -- (BB2);
        \node[right=0.13 of Z, xshift=0.8 cm, yshift=0.8 cm] (QQn) {};
        \node[right=1.68 of Z, xshift=0.8 cm, yshift=0.8 cm] (BBn) {};
        \draw[->] (QQn) -- (BBn);
        \node[draw, rectangle, fit=(S), inner sep=1pt, minimum height=2 cm] (RectangleS) {};
        \node[left=0.35 of Z] (ZSS) {};
        \node[draw, dashed, fit=(RectangleS)(ZSS), inner sep=5pt] (SS) {};
        \node[above=0.1cm of SS, font=\small] {Semantic Source};
\end{tikzpicture}}
    \caption{Probabilistic graphical modeling of the proposed semantic source \cite{10654356}.}
    \label{fig:semantic_source_model}
\end{figure}
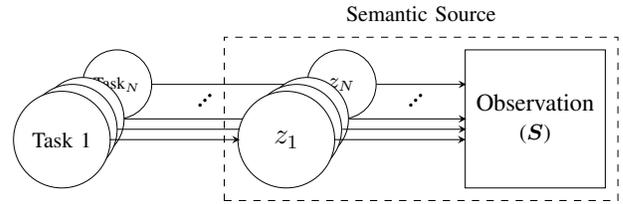

Fig. \ref{fig:semantic_source_model} illustrates our interpretation of \emph{semantic source} as discussed in \cite{10654356}. Such a definition enables the simultaneous extraction of multiple \emph{semantic variables} based on a single observation and addresses multiple tasks. It consists of $N$ semantic variables, denoted by $\bm{z}=[\,z_1\, z_2\,\dots\,z_N]\,$, lying behind an observation $\bm{S}$, and the given tasks specify one/multiple semantics to be of our interest. The tuple of $(\bm{z}, \bm{S})$ is defined as the semantic source, fully described by the probability distribution of $p(\bm{z}, \bm{S})$. In this study, we focus on a more practical scenario of the distributed setting, where instead of the full observation, multiple partial views of the data are available. These partial observations, denoted as $\bm{S_1}, \dots\,,\bm{S_K}$, each contain information about some or all semantic variables.

In \cite{10654356}, it was demonstrated that when semantic variables share statistical relationships, a split semantic encoder, comprising a CU and multiple SUs, enables cooperative SemCom, significantly improving performance in multi-task cases by utilizing common information. In realistic scenarios, sensing nodes only access partial observations of the source, and collaborative approaches are required to perform tasks based on these distributed views. 

\begin{figure}[!t]
\centering
\scalebox{0.99}{
\begin{tikzpicture}
    \node[draw, rectangle, inner sep=0.5pt] (Sa) {
        $\begin{tabular}{@{}c@{}}
        \text{Semantic} \\
        \text{Source} \\
        \text{$(\bm{z}, \bm{S})$}
        \end{tabular}$
    };

    \node[below=2.5 cm of Sa.west] (S) {};
    
    \node[draw, rectangle, right=0.35 cm of S, yshift=1.5cm, inner sep=2pt] (PO1) {obs. $\bm{S}_1$};
    \node[draw, rectangle, right=0.35 cm of S, yshift=-1.0cm, inner sep=2pt] (PO4) {obs. $\bm{S}_4$};

    \draw[->, line width=1pt, shorten >=2pt, shorten <=2pt] ($ (Sa.south) + (-0.4cm,0) $) -- ($ (PO1.west) + (-0.20cm,0) $) -- ($ (PO1.west) + (+0.05cm,0) $);
    \draw[->, line width=1pt, shorten >=2pt, shorten <=2pt] ($ (Sa.south) + (-0.5cm,0) $)-- ($ (PO4.west) + (-0.30cm,0) $) -- ($ (PO4.west) + (+0.05cm,0) $);
    
    \node[draw, rectangle, right=0.4 cm of PO1, inner sep=2pt] (CU1) {$\text{CU}1$};
    \node[draw, rectangle, right=0.4 cm of PO4, inner sep=2pt] (CU4) {$\text{CU}4$};
    
    \draw[->, line width=1pt, shorten >=2pt, shorten <=2pt] (PO1.east) -- (CU1.west);
    \draw[->, line width=1pt, shorten >=2pt, shorten <=2pt] (PO4.east) -- (CU4.west);

    \node[draw, rectangle, right=0.4 cm of CU1, yshift=0.4cm, inner sep=2pt] (SU11) {$\begin{tabular}{@{}c@{}}
        \text{SU $1.1$} 
        \end{tabular}$};
    \node[draw, rectangle, right=0.4 cm of CU1, yshift=-0.4cm, inner sep=2pt] (SU12) {$\begin{tabular}{@{}c@{}}
        \text{SU $1.2$} 
        \end{tabular}$};
    \draw[->, line width=1pt, shorten >=2pt, shorten <=2pt] (CU1.east) -- ($ (CU1.east) + (0.15,0)$) -- ($(SU11.west) + (-0.25,0)$) -- (SU11.west);
    \draw[->, line width=1pt, shorten >=2pt, shorten <=2pt] (CU1.east) -- ($ (CU1.east) + (0.15,0)$) -- ($(SU12.west) + (-0.25,0)$) -- (SU12.west);

    \node[draw, rectangle, right=0.4 cm of CU4, yshift=0.4cm, inner sep=2pt] (SU41) {$\begin{tabular}{@{}c@{}}
        \text{SU $4.1$} 
        \end{tabular}$};
    \node[draw, rectangle, right=0.4 cm of CU4, yshift=-0.4cm, inner sep=2pt] (SU42) {$\begin{tabular}{@{}c@{}}
        \text{SU $4.2$} 
        \end{tabular}$};
    \draw[->, line width=1pt, shorten >=2pt, shorten <=2pt] (CU4.east) -- ($ (CU4.east) + (0.15,0)$) -- ($(SU41.west) + (-0.25,0)$) -- (SU41.west);
    \draw[->, line width=1pt, shorten >=2pt, shorten <=2pt] (CU4.east) -- ($ (CU4.east) + (0.15,0)$) -- ($(SU42.west) + (-0.25,0)$) --(SU42.west);

    \node[draw, dashed, fit=(CU1) (SU11) (SU12), inner sep=1pt] {};
    \node[above=0.40 cm of CU1,xshift=0.6cm] (SN1) {Sensing node $1$};

    \node[draw, dashed, fit=(CU4) (SU41) (SU42), inner sep=1pt] {};
    \node[above=0.40 cm of CU4,xshift=0.6cm] (SN4) {Sensing node $4$};

    \draw[loosely dotted, line width=1pt] ($ (PO1) - (0,0.42cm) $) -- ($ (PO4) + (0,0.42cm) $);

    \draw[loosely dotted, line width=1pt] ($ (SN1) - (0,1.75cm) $) -- ($ (SN4) + (0,0.30cm) $);

    \node[draw, rectangle, right=2.7 cm of SU11, inner sep=2pt,yshift = -0.45cm] (Dec1) {$\begin{tabular}{@{}c@{}}
        \text{Decoder} \\
        \text{Task $1$}
        \end{tabular}$};
    \node[draw, rectangle, right=2.7 cm of SU42, inner sep=2pt, yshift = +0.45cm] (Dec2) {$\begin{tabular}{@{}c@{}}
        \text{Decoder} \\
        \text{Task $2$}
        \end{tabular}$};
    
    \draw[->, line width=1pt, shorten >=2pt, shorten <=2pt] (SU11.east) -- ($ (Dec1.west) + (0,+0.45cm) $);
    \draw[->, line width=1pt, shorten >=2pt, shorten <=2pt] ($ (Dec1.west) + (-2.00,+0.15cm) $) -- ($ (Dec1.west) + (0,+0.15cm) $);
    \node[left=1.9 cm of Dec1, inner sep=2pt,yshift = 0.15cm] (CU2c) {$\text{...}$};
    \draw[->, line width=1pt, shorten >=2pt, shorten <=2pt] ($ (Dec1.west) + (-2.00,-0.15cm) $) -- ($ (Dec1.west) + (0,-0.15cm) $);
    \node[left=1.9 cm of Dec1, inner sep=2pt,yshift = -0.15cm] (CU3c) {$\text{...}$};
    \draw[->, line width=1pt, shorten >=2pt, shorten <=2pt] (SU41.east)  -- ($ (Dec1.west) + (-2.0,-0.45) $) -- ($ (Dec1.west) + (0,-0.45) $);
    
    \draw[->, line width=1pt, shorten >=2pt, shorten <=2pt] (SU12.east)  -- ($ (Dec2.west) + (-2.0,+0.45) $) -- ($ (Dec2.west) + (0,+0.45) $);
    \draw[->, line width=1pt, shorten >=2pt, shorten <=2pt] ($ (Dec2.west) + (-2.0,+0.15cm) $) -- ($ (Dec2.west) + (0,+0.15cm) $);
    \node[left=1.9 cm of Dec2, inner sep=2pt,yshift = 0.15cm] (CU2c2) {$\text{...}$};
    \draw[->, line width=1pt, shorten >=2pt, shorten <=2pt] ($ (Dec2.west) + (-2.0,-0.15cm) $) -- ($ (Dec2.west) + (0,-0.15cm) $);
    \node[left=1.9 cm of Dec2, inner sep=2pt,yshift = -0.15cm] (CU3c2) {$\text{...}$};
    \draw[->, line width=1pt, shorten >=2pt, shorten <=2pt] (SU42.east) -- ($ (Dec2.west) + (0,-0.45) $);

    \node[draw,right=2.6 cm of CU1, inner sep=1pt, rectangle, yshift=-0.0cm,fill=white] (ch) {$\begin{tabular}{@{}c@{}}
          \tiny \\
          \scriptsize Channel $1$ \\
          \tiny
    \end{tabular}$};
    \node[draw,right=2.6 cm of CU4, inner sep=1pt, rectangle, yshift=+0.0cm,fill=white] (ch) {$\begin{tabular}{@{}c@{}}
          \tiny \\
          \scriptsize Channel $2$ \\
          \tiny
    \end{tabular}$};
    
\end{tikzpicture}
}
\caption{An illustration of the proposed CCMT system model for distributed partial observations for $N\!=\!2$ and $K\!=\!4$.}
    \label{fig:system-model}
\end{figure}
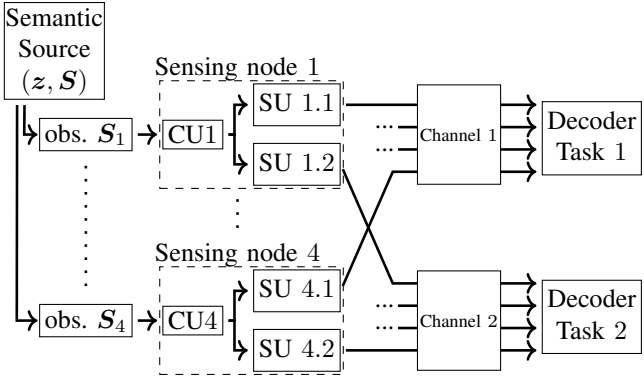

To integrate cooperative multi-tasking with collaborative handling of distributed data, our proposed system model consists of $K$ observations available at $K$ sensing nodes and $N$ semantic variables, each associated with a unique task. As illustrated in Fig. \ref{fig:system-model}, at each sensing node, the CU encoder first extracts the common relevant information from its observation. Next, $N\times K$ SU encoders extract and transmit task-specific information to their respective decoders for collaborative decoding. To fulfill each task, the corresponding receivers need $K$ transmitted information extracted and transmitted by assigned SUs from each observation. Since we consider the execution of two tasks, two SUs are required at each sensing node. Thus, we show the output of SU encoders as $\bm{x}_{1.1},\, \bm{x}_{1.2},\dots,\bm{x}_{K.1},\, \bm{x}_{K.2}$, and their noise-corrupted version received at the corresponding decoders are indicated by $\bm{\hat{x}}_{1.1},\, \bm{\hat{x}}_{1.2},\dots,\bm{\hat{x}}_{K.1},\, \bm{\hat{x}}_{K.1}$.

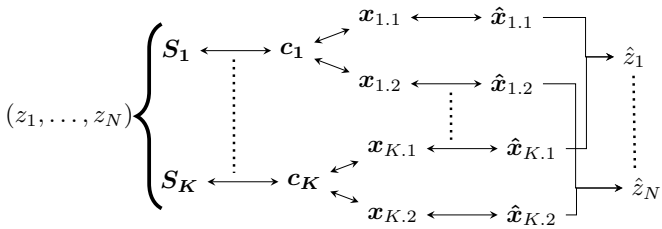
\begin{figure}[H]
    \centering
    \resizebox{0.5\textwidth}{!}{\begin{tikzpicture}[>=stealth]

    \node[above=0.8cm of {0,0}] (Z1) {};
    \node[below=0.8cm of {0,0}] (Zn) {};
    \node[right=0.3cm of Z1] (S1) {$\bm{S_1}$};
    \node[right=0.28cm of Zn] (Sn) {$\bm{S_K}$};
    \node[right=of S1] (C1) {$\bm{c_1}$};
    \node[right=of Sn] (Cn) {$\bm{c_K}$};
    \node[above right=0.01cm and 0.5cm of C1] (X11) {$\bm{x}_{1.1}$};
    \node[below right=0.01cm and 0.5cm of C1] (X12) {$\bm{x}_{1.2}$};
    \node[right=of X11] (X1h) {$\bm{\hat{x}}_{1.1}$};
    \node[right=of X12] (X2h) {$\bm{\hat{x}}_{1.2}$};
    \node[above right=0.01cm and 0.4cm of Cn] (X41) {$\bm{x}_{K.1}$};
    \node[below right=0.01cm and 0.4cm of Cn] (X42) {$\bm{x}_{K.2}$};
    \node[right=of X41] (X41h) {$\bm{\hat{x}}_{K.1}$};
    \node[right=of X42] (X42h) {$\bm{\hat{x}}_{K.2}$};

    \node[right=0.35cm of S1] (aux1) {};
    \node[right=0.25cm of Sn] (aux2) {};

    \node[right=0.44cm of X12] (aux3) {};
    \node[right=0.22cm of X41] (aux4) {};

    \draw[dotted, line width=1pt] (aux1) -- (aux2);
    \draw[dotted, line width=1pt] (aux3) -- (aux4);
    \draw[<->] (S1) -- (C1);
    \draw[<->] (Sn) -- (Cn);
    \draw[<->] (C1) -- (X11);
    \draw[<->] (C1) -- (X12);
    \draw[<->] (Cn) -- (X41);
    \draw[<->] (Cn) -- (X42);
    
    \draw[<->] (X11) -- (X1h);
    \draw[<->] (X12) -- (X2h);
    \draw[<->] (X41) -- (X41h);
    \draw[<->] (X42) -- (X42h);    

    \draw[line width=1.5pt,decorate,decoration={brace,amplitude=10pt},xshift=0.4cm] ({0.15,-1.3}) -- ({0.15,1.3}) node[midway,xshift=-1.3cm]{$(z_1, \dots, z_N)$};

    \node[below right=0.01cm and 1cm of X1h] (Z1h) {$\hat{z}_1$}; 
    \draw[->] (X1h.east) -- ++(0.6cm,0) |- (Z1h.west); 
    \draw[->] (X41h.east) -- ++(0.29cm,0) |- (Z1h.west);

    \node[below right=0.01cm and 0.8cm of X41h] (Znh) {$\hat{z}_N$};
    \draw[->] (X2h.east) -- ++(0.46cm,0) |- (Znh.west); 
    \draw[->] (X42h.east) -- ++(0.15cm,0) |- (Znh.west);

    \node[below right=0.3cm and 1.15cm of X1h] (aux5) {};
    \node[below right=0.01cm and 0.85cm of X41h] (aux6) {};
    \draw[dotted, line width=1pt] (aux5) -- (aux6);

\end{tikzpicture}}
    \caption{Markov representation of the CCMT system.}
    \label{fig:markov_chain}
\end{figure}

Our approach incorporates wireless channels between encoders and decoders, employing the additive white Gaussian noise (AWGN) channel. As shown generally in Fig. \ref{fig:markov_chain}, the Markov representation of our system model for the $i$-th semantic variable is outlined for $\forall k \in \{1,\dots,K\}$ as follows.
\begin{equation}
\begin{split}
    &p(\hat{z}_i,\bm{\hat{x}}_{k.i},\bm{x}_{k.i},\bm{c}_{k}|\bm{S}_{k}) =\\[0.5em] & p^{\text{\tiny Dec$_i$}}(\hat{z}_i|\bm{\hat{x}}_{k.i})\,p^{\text{\tiny Channel$_k$}}(\bm{\hat{x}}_{k.i}|\bm{x}_{k.i})\,p^{\text{\tiny SU$_{k.i}$}}(\bm{x}_{k.i}|\bm{c}_{k})\,p^{\text{\tiny CU$_k$}}(\bm{c}_k|\bm{S}_k).
\end{split}
\label{eq:system_probability}
\end{equation}
In (\ref{eq:system_probability}), $p^{\text{\tiny CU$_k$}}(\bm{c}_k|\bm{S}_k)$ defines the CU of the $k$-th sensing node, which extracts the common relevant information available in the $k$-th observation amongst all tasks. The corresponding SU for $i$-th semantic variable at $k$-th sensing node is described by $p^{\text{\tiny SU$_{k.i}$}}(\bm{x}_{k.i}|\bm{c}_k)$ extracting task-specific information and providing $\bm{x}_{k.i}$ as the channel input. The corresponding decoder is then specified by $p^{\text{\tiny Dec$_i$}}(\hat{z}_i|\bm{\hat{x}}_{k.i})$, where $\bm{\hat{x}}_{k.i} \in \mathbb{R}^{m_i}$, $\forall k \in \{1,\dots,K\}$ is the received information passed through the AWGN channel and modeled like $\bm{\hat{x}}_{k.i} = \bm{x}_{k.i} + \bm{n}$, where $\bm{n} \sim \mathcal{N}(\mathbf{0}_{m_i}, \sigma^2_{i} \mathbf{I}_{m_i})$, and $m_i$ is the size of the encoded task-specific information or the number of channel uses.
\subsection{Optimization Problem} \label{subsec.Loss_function}
We formulate an optimization problem by adopting the information maximization principle together with the E2E learning method, as follows.
\begin{equation} \label{eq:optimization_problem}
\begin{split}
    &\left[\,p^{\text{\tiny CU$_k$}}(\bm{c}_k|\bm{S}_k)^\star,\, p^{\text{\tiny SU$_k$}}(\bm{x}_{k}|\bm{c}_k)^\star\right]\ _{k=1}^{K} = \\[0.6em] & \qquad \qquad \qquad \qquad \arg \underset{\substack{p^{\text{\tiny CU$_k$}}(\bm{c}_k|\bm{S}_k), \\ p^{\text{\tiny SU$_k$}}(\bm{x}_k|\bm{c}_k).}}{\text{max}}\, \sum_{i=1}^{N}\, I(z_i ; \bm{\hat{x}}_{(1:K).i}).
\end{split}
\end{equation}

Thus, the objective is to maximize the mutual information between the channel outputs $\bm{\hat{x}}_{(1:K).i}$ of the corresponding SUs, and the semantic variables $z_i$. Expanding the mutual information in (\ref{eq:optimization_problem}) as discussed in detail in \cite{10654356}, the approximated objective function is derived like:
\begin{equation} \label{eq:objective_function}
    \begin{aligned}
        &\mathcal{L}^{\text{\tiny CCMT}}(\boldsymbol{\theta}, \boldsymbol{\Phi}) = \sum_{i=1}^{N}\, I(z_i ; \bm{\hat{x}}_{(1:K).i}) \\[0.6em]
        &\approx \mathbb{E}_{p^{\text{\tiny CU}}_{\boldsymbol{\theta}}(\bm{c}_{1:K}|\bm{S}_{1:K})}\left[\,\sum_{i=1}^{N} \left\{\mathbb{E}_{p(\bm{S}_{1:K},z_i)} \bigg[\, \right. \right. \\[0.6em]
        & \qquad \qquad \qquad \left. \left. \mathbb{E}_{p^{\text{\tiny SU}}_{\boldsymbol{\Phi}}(\bm{\hat{x}}_{(1:K).i}|\bm{c}_{1:K})} [\,\log p(z_i|\bm{\hat{x}}_{(1:K).i})]\ \bigg]\,\right\} \right]\,.
    \end{aligned}
\end{equation}

To derive the objective function on (\ref{eq:objective_function}), we have employed the variational method, which is a way to approximate intractable computations based on some adjustable parameters, like weights in NNs \cite{kingma2013auto}. The technique is widely used in machine learning, e.g., \cite{alemi2016deep}, and also in task-oriented communications, e.g., \cite{Shao2021}, \cite{Shao2022}, and \cite{Beck2023}. Thus, our posterior distributions, $\{p^{\text{\tiny CU$_k$}}(\bm{c}_k|\bm{S}_k)\}_{k=1}^{K}$ and $\{p^{\text{\tiny SU$_k$}}(\bm{\hat{x}}_k|\bm{c}_k)\}_{k=1}^{K}$ are approximated by NN parameters of $\boldsymbol{\theta} = \{ \theta_k \}_{k=1}^K$ and $\boldsymbol{\Phi} = \{ \phi_k \}_{k=1}^K$ respectively.

As shown in (\ref{eq:objective_function}), by considering the channel outputs we aim to emphasize the role of joint semantic and channel coding performed by our SUs. Employing the fact that $p^{\text{\tiny SU$_k$}}_{\boldsymbol{\phi}_k}(\bm{\hat{x}}_{k}|\bm{c}_k)=\int p^{\text{\tiny SU$_k$}}_{\boldsymbol{\phi}_k}(\bm{x}_{k}|\bm{c}_k)\,p^{\text{\tiny Channel$_k$}}(\bm{\hat{x}}_k|\bm{x}_k)\,d\bm{x}_k$, we try to optimize $p^{\text{\tiny SU$k$}}_{\boldsymbol{\phi}_k}(\bm{\hat{x}}_k|\bm{c}_k)$.

Regarding the $i$-th decoder in (\ref{eq:objective_function}), the $p^{\text{\tiny Dec$_i$}}(z_i|\bm{\hat{x}}_{(1:K).i})$ can be fully determined using the known distributions and underlying probabilistic relationship in (\ref{eq:system_probability}) as:
\begin{equation} \label{eq:decoder_probability}
    p^{\text{\tiny Dec$_i$}}(z_i|\bm{\hat{x}}_{k.i}) = \frac{\int p^{\text{\tiny SU$_k$}}_{\phi_k}(\bm{\hat{x}}_{k.i}|\bm{c}_{k})\,p^{\text{\tiny CU$_k$}}_{\theta_k}(\bm{c}_{k}|\bm{S}_k)\,p(\bm{S}_k,z_i)\,d\bm{s}_k\,d\bm{c}_k}{p(\bm{\hat{x}}_{k.i})},
\end{equation}
however, due to the high-dimensional integrals, (\ref{eq:decoder_probability}) becomes intractable and we need to follow the variational approximation technique, resulting in the following:
\begin{equation} \label{eq:objective_function_variational}
    \begin{aligned}
        &\mathcal{L}^{\text{\tiny CCMT}}_{\text{\tiny approx.}}(\boldsymbol{\theta}, \boldsymbol{\Phi}, \boldsymbol{\Psi}) = \sum_{i=1}^{N}\, I(z_i ; \bm{\hat{x}}_{(1:K).i}) \\[0.6em]
        &\approx \mathbb{E}_{p^{\text{\tiny CU}}_{\boldsymbol{\theta}}(\bm{c}_{1:K}|\bm{S}_{1:K})}\left[\,\sum_{i=1}^{N} \left\{\mathbb{E}_{p(\bm{S}_{1:K},z_i)} \bigg[\, \right. \right. \\[0.6em]
        & \qquad \qquad \qquad \left. \left. \mathbb{E}_{p^{\text{\tiny SU}}_{\boldsymbol{\Phi}}(\bm{\hat{x}}_{(1:K).i}|\bm{c}_{1:K})} [\,\log p^{\text{\tiny Dec$_i$}}_{\boldsymbol{\psi}_i}(z_i|\bm{\hat{x}}_{(1:K).i})]\ \bigg]\,\right\} \right]\,.
    \end{aligned}
\end{equation}

Where in (\ref{eq:objective_function_variational}), $\boldsymbol{\Psi} = \{ \psi_i \}_{i=1}^N$ represents NN parameters approximating the true distribution of decoders. To obtain the empirical estimate of the above objective function, we approximate the expectations using Monte Carlo sampling assuming the existence of a dataset $\{\mathbf{S}^{(j)}, z^{(j)}_1,\dots,z^{(j)}_N\}^{J}_{j=1}$ where $J$ represents the batch size of the dataset \cite{bishop2006pattern}.
\begin{equation} \label{eq:empirical_objective}
    \begin{aligned}
        &\mathcal{L}^{\text{\tiny CCMT}}_{\text{\tiny empir.}}(\boldsymbol{\theta}, \boldsymbol{\Phi}, \boldsymbol{\Psi}) \approx \\[0.6em] &\frac{1}{L}\sum_{l=1}^{L}\left[\,\sum_{i=1}^{N}\left\{\frac{1}{J}\sum_{j=1}^{J}\bigg[\,\frac{1}{T}\sum_{t=1}^{T} [\,\log q^{\text{\tiny Dec$_i$}}_{\boldsymbol{\psi}_i}(z_i|\bm{\hat{x}}_{(1:K).j,t})]\,\bigg]\,\right\}\right]\,.
    \end{aligned}
\end{equation}

In (\ref{eq:empirical_objective}), $L$ represents the sample size of the cooperative processing, and $T$ is the channel sampling size for each batch.  

\section{Simulation results}\label{sec:simulation_results}

For the distributed scenario, we compare our proposed CCMT with a baseline approach called single-task collaborative (STC) semantic communication with no cooperative multitask processing. We evaluate the task execution error rate for task $1$ and task $2$ across different training scenarios, channel conditions, and NN sizes, demonstrating the performance improvements of CCMT over STC\footnote{The code is available at https://github.com/ahmadhalimi95/CCMT}.

\subsection{Simulation Setup}
We evaluate our proposed architecture for four sensing nodes that need to collaborate and cooperate on two tasks, where each sensing node has a partial observation of exactly one quarter of the full image showing a digit of the MNIST dataset \cite{deng2012mnist}. For our simulations, we consider task $1$ to be the binary classification of digit two and task $2$, the categorical classification of the digits in the MNIST dataset. To examine more practical and complex scenarios, in which sensing nodes may view the observing object from different angles, we consider situations where each quarter image is individually rotated by for a randomly selected angle.

For the semantic encoders we use convolutional NNs (CNNs) with ReLU activation functions and max-pool layers for image size reduction, as specified in Tab. \ref{table_implementation_distributed}. 
In our simulations, we fix the CU to two CNN layers and the SU to one CNN layer with a fully connected (FC) layer for each task. In the STC case, where no CU is used, the SUs consist of three CNN layers and one FC layer each, to make the number of layers equivalent to the CCMT case for a fair comparison.
The number of filters for each CNN layer in CCMT: $c_1,c_2,c_3$ and STC: $k_1,k_2,k_3$, is set for each comparison such that the total number of parameters in CCMT and STC are approximately equal. 

For all simulations, the number of channel uses for task $1$ and task $2$ is two per sensing node, i.e., $m_1=m_2=2$. For Figs. \ref{fig:over_epochs} and \ref{fig:over_SNR_both}, 
$c_1=6,c_2=5,c_3=3$, and $k_1=4,$ $k_2=4,k_3=3$ are used, which means for the CNN layers of the CCMT architecture a total of $(9+1)c_1 + (9c_1+1)c_2 + 2((9c_2+1)c_3) = 611$ parameters and for the STC architecture $598$ parameters are used. The final layer of each encoder normalizes the output power across the channel uses to average the output power for each transmission to one over each NN training batch. The signal-to-noise ratio (SNR) of the AWGN channel is defined as $\text{SNR} := 1/\sigma^2_i$, where $\sigma^2_i$ is the noise power of the zero mean i.i.d. Gaussian noise vector of each channel $\bm{n} \in \mathbb{R}^{m_i}$. For the simulations, \num{60000} training and \num{10000} validation data samples are used, the results are shown for the validation dataset, and the results of all simulations are averaged over $25$ independent iterations. 
\begin{table}[!t]
    \centering
    \renewcommand{\arraystretch}{1.0}
    \setlength{\tabcolsep}{5pt}
    \caption{The NN structure for each distributed sensing node}
    \begin{tabular}{p{42mm}|p{15mm}|p{19mm}}
        \hline
        \multicolumn{3}{p{82mm}}{All CNN layers are $3\times3$ convolution filters with stride $1$ and zero padding to keep the input and output dimensions the same. As the MNIST dataset consists of black and white $28$ by $28$ pixels, the input images for each of the four agents are $14$ by $14$ pixels. The learning rate is set to $10^{-4}$, but for the last $30$ epochs reduced by a factor of $0.1$ per ten epochs. For the case with image rotation without CU, the learning rate is reduced by a factor of $0.1$ for the last $200$, last $100$, and last $50$ epochs each.}\\
        \hline
           & \textbf{Output size} & \textbf{No. $\!$ of $\!$ param.}  \\
        \hline
        \hline
        
        \multicolumn{3}{p{82mm}}{\textbf{Encoder Layers for CCMT (for task $1$ and $2$)} }    \\
        \hline
        CU: CNN layer, ReLU, max-pool \newline
         CU: CNN layer, ReLU, max-pool & $7\times7\times c_1$ \newline 
         $3 \times 3 \times   c_2$
         & $(9 +1) c_1$ \newline $(9c_1+1) c_2$ \\
        \hline
         $\text{SU}_{\text{task} 1} $:$\!$ CNN layer, ReLU \newline 
         $\text{SU}_{\text{task} 1} $:$\!$ FC, power normalization   \newline
         $\text{SU}_{\text{task} 2} $:$\!$ CNN layer, ReLU \newline
         $\text{SU}_{\text{task} 2} $:$\!$ FC, power normalization & $3\times3\times c_3$ \newline 
         $m_1$ \newline $3\times3\times c_3$ \newline $m_2$ & $(9c_2+1) c_3$ \newline $(9c_3+1)m_1$ \newline $(9c_2+1) c_3$ \newline $(9c_3+1)m_2$ \\
        \hline
        \hline
        \multicolumn{3}{p{82mm}}{\textbf{Encoder Layers for STC} (for task $i=1,2$ ) } \\
        \hline
        $\text{SU}_{\text{task} i} $:$\!$ CNN layer, ReLU, max-pool \newline
        $\text{SU}_{\text{task} i} $:$\!$ CNN layer, ReLU, max-pool \newline
        $\text{SU}_{\text{task} i} $: $\!$ CNN layer, ReLU, \newline
        $\text{SU}_{\text{task} i}$: $\!$ FC, power normalization
        & $7\times7\times k_1$ \newline 
        $3\times3\times k_2$ \newline 
        $3\times3\times k_3$ \newline
        $m_i$ & 
        $(9+1)k_1$ \newline
        $(9k_1+1) k_2$ \newline
        $(9k_2+1) k_3$ \newline
        $(9k_3+1) m_i$
        \\
        \hline
        \hline
        \multicolumn{3}{p{82mm}}{\textbf{Decoder} }    \\
        \hline
        Decoder task $1$: FC, Tanh \newline
        Decoder task $1$: FC, Sigmoid \newline
        Decoder task $2$: FC, Tanh \newline
        Decoder task $2$: FC, Softmax & 
        $16$ \newline
        $1$ \newline
        $16$ \newline
        $10$ &
        $(m_1+1) 16$ \newline
         $ 16+1 $ \newline
        $(m_2+1)  16 $ \newline
        $(16+1)10$  \\
        \hline
    \end{tabular}
    \label{table_implementation_distributed}
\end{table}

\subsection{Training Scenarios}
\begin{figure}[t!]
	\centering
    \vspace{-5.58cm} 
    \textcolor{white}{.}
    \resizebox{0.44\textwidth}{!}{\definecolor{brown1363485}{RGB}{136,34,85}
\definecolor{burlywood221204119}{RGB}{221,204,119}
\definecolor{darkgray176}{RGB}{176,176,176}
\definecolor{darkslateblue5134136}{RGB}{51,34,136}
\definecolor{forestgreen1711951}{RGB}{17,119,51}
\definecolor{lightgray204}{RGB}{204,204,204}

\begin{tikzpicture}
\begin{axis}[ %
 hide axis, 
xmin = 10, 
xmax = 50, 
ymin = 0, 
ymax = 0.4, 
legend columns = 1, 
legend style={at={(0.25,0)}, 
legend style={font=\footnotesize},
anchor = north west, 
draw = white!15!black, 
legend cell align = left}] 

\addlegendimage{thick, color=brown1363485}
\addlegendentry{STC}


\addlegendimage{thick, color=burlywood221204119}
\addlegendentry{CCMT}

\addlegendimage{dashed, thick, color=darkslateblue5134136}
\addlegendentry{CCMT-generalized-CU} 

\end{axis} 
\end{tikzpicture}}
    \vspace{0.1cm}
	\textcolor{white}{.}
    \resizebox{0.44\textwidth}{!}{\input{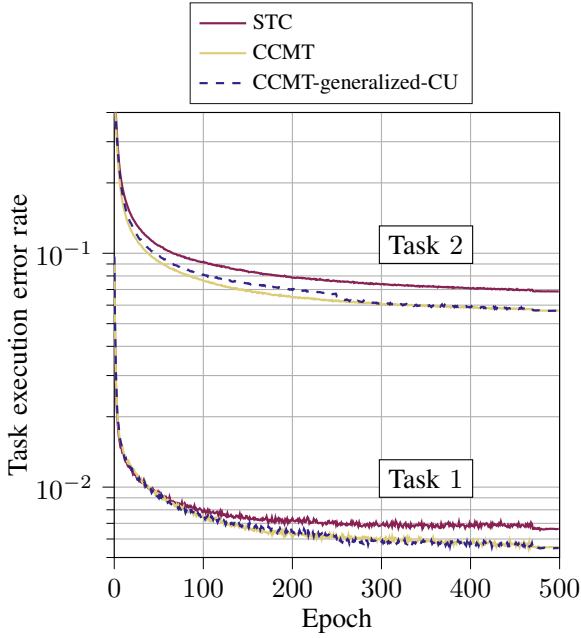}}
	\caption{Task execution error rate over the number of training epochs of task $1$ and task $2$ for different training scenarios.}
    \label{fig:over_epochs}
\end{figure}

In Fig. \ref{fig:over_epochs} the task execution error rate of task $1$ and task $2$ over the number of epochs for different SNRs is shown. It is worth mentioning that the SNR is uniformly distributed for all ranges, and specifically for Fig. \ref{fig:over_epochs}, the evaluation SNR range is the same as the training SNR range. In this figure, the proposed CCMT architecture is compared to the STC architecture when both are trained for $500$ epochs for an SNR range from $9$ to $11 \, \text{dB}$. It can be seen that the CCMT outperforms the STC in task execution error rate.

To further investigate the joint semantic and channel coding performance of the SUs to deal with different channel conditions, the CCMT is first trained for a wide SNR range from $-10$ to $20 \, \text{dB}$ for $250$ epochs, causing the CU to be generalized and then the CU is frozen and the SUs are further trained for the smaller target SNR range of $9$ to $11 \, \text{dB}$ for additional $250$ epochs. It can be seen in Fig. \ref{fig:over_epochs} that this approach, named ``CCMT-generalized-CU", performs equally, or even slightly better than the case where the whole CCMT is trained for a small SNR range. The validation SNR for all results in Fig. \ref{fig:over_epochs} is set to $10\, \text{dB}$. 

We conclude that the generalization of the CU has the advantage that only retraining of the SUs is required to deal with different channel conditions for task $1$ and task $2$. Therefore, we use the CCMT-generalized-CU, where the CU is trained for $250$ epochs for an SNR range of $-10$ to $20 \, \text{dB}$, and then the SUs are trained for specific SNR ranges depending on the channel conditions.

\subsection{Impact of Different Channel Conditions}
\begin{figure}[t!]
    \centering
    \vspace{-6.6cm} 
    \textcolor{white}{.}
    \resizebox{0.50\textwidth}{!}{\definecolor{brown1363485}{RGB}{136,34,85}
\definecolor{cadetblue68170153}{RGB}{68,170,153}
\definecolor{darkorchid17068153}{RGB}{170,68,153}
\definecolor{darkslateblue5134136}{RGB}{51,34,136}

\begin{tikzpicture}
\begin{axis}[ %
 hide axis, 
xmin = 10, 
xmax = 50, 
ymin = 0, 
ymax = 0.4, 
legend columns = 1, 
legend style={at={(0,0)}, 
legend style={font=\scriptsize},
anchor = north west, 
draw = white!15!black, 
legend cell align = left}] 

\addlegendimage{thick, color=darkorchid17068153, mark=triangle, mark size=1.8, mark options={solid,fill=none}}
\addlegendentry{STC: single-model}

\addlegendimage{thick, color=cadetblue68170153, mark=square, mark size=1.8, mark options={solid,fill=none}}
\addlegendentry{CCMT: single-model}

\addlegendimage{thick, color=brown1363485, mark=diamond, mark size=1.8, mark options={solid,fill=none}}
\addlegendentry{STC: multi-model (best selected)}

\addlegendimage{thick, color=darkslateblue5134136, mark=o, mark size=1.8, mark options={solid,fill=none}}
\addlegendentry{CCMT-generalized-CU: multi-model  (best selected)}

\end{axis} 
\end{tikzpicture}}
    \vspace{-0.5cm}
    \textcolor{white}{.}
    
    \begin{subfigure}{0.44\textwidth} 
        \centering
        \resizebox{\textwidth}{!}{
\begin{tikzpicture}
\pgfplotsset{%
    width=0.9\textwidth,
    height=0.95\textwidth 
}   
\definecolor{brown1363485}{RGB}{136,34,85}
\definecolor{cadetblue68170153}{RGB}{68,170,153}
\definecolor{darkgray176}{RGB}{176,176,176}
\definecolor{darkorchid17068153}{RGB}{170,68,153}
\definecolor{darkslateblue5134136}{RGB}{51,34,136}
\definecolor{lightgray204}{RGB}{204,204,204}

\begin{axis}[
legend cell align={left},
legend style={fill opacity=0.8, draw opacity=1, text opacity=1, draw=lightgray204},
log basis y={10},
minor xtick={},
minor ytick={0.0002,0.0003,0.0004,0.0005,0.0006,0.0007,0.0008,0.0009,0.002,0.003,0.004,0.005,0.006,0.007,0.008,0.009,0.02,0.03,0.04,0.05,0.06,0.07,0.08,0.09,0.2,0.3,0.4,0.5,0.6,0.7,0.8,0.9,2,3,4,5,6,7,8,9,20,30,40,50,60,70,80,90},
tick align=outside,
tick pos=left,
x grid style={darkgray176},
xlabel={SNR},
xmajorgrids,
xmin=-10, xmax=20,
xtick style={color=black},
xtick={-15,-10,-5,0,5,10,15,20,25},
y grid style={darkgray176},
ylabel={Task execution error rate},
ymajorgrids,
yminorgrids,
ymin=0.004, ymax=1,
ymode=log,
ytick style={color=black},
ytick={0.0001,0.001,0.01,0.1,1,10},
yticklabels={
  \(\displaystyle {10^{-4}}\),
  \(\displaystyle {10^{-3}}\),
  \(\displaystyle {10^{-2}}\),
  \(\displaystyle {10^{-1}}\),
  \(\displaystyle {10^{0}}\),
  \(\displaystyle {10^{1}}\)
}
]
\node[draw=black, fill=white, rectangle, anchor=south] at (axis cs:11.5,0.12) {Task $2$};
\node[draw=black, fill=white, rectangle, anchor=south] at (axis cs:-4,0.012) {Task $1$};
\addplot [ thick, darkorchid17068153, mark=triangle, mark size=1.5, mark options={solid,fill=none}]
table {%
-12 0.183144018054008
-11 0.168376043438911
-10 0.15242400765419
-9 0.136520013213158
-8 0.121020033955574
-7 0.105708032846451
-6 0.0904040262103081
-5 0.0754080191254616
-4 0.0622560158371925
-3 0.0502280220389366
-2 0.0399200282990932
-1 0.0312320254743099
0 0.0244720242917538
1 0.0194680206477642
2 0.0160480197519064
3 0.0137720247730613
4 0.0119000244885683
5 0.0107240295037627
6 0.00973602570593357
7 0.00914402678608894
8 0.0086080264300108
9 0.00828001741319895
10 0.00801202282309532
11 0.00774401659145951
12 0.00754002807661891
13 0.00738402828574181
14 0.00724002858623862
15 0.00716002471745014
16 0.00710802571848035
17 0.0070400214754045
18 0.00700402725487947
19 0.00695602176710963
20 0.00694002630189061
};
\addplot [ thick, cadetblue68170153, mark=square, mark size=1.5, mark options={solid,fill=none}]
table {%
-12 0.183564037084579
-11 0.167688027024269
-10 0.150720030069351
-9 0.134272038936615
-8 0.11811201274395
-7 0.102112010121346
-6 0.0869680196046829
-5 0.0722600221633911
-4 0.0592120178043842
-3 0.0473720170557499
-2 0.0371360220015049
-1 0.0290480274707079
0 0.0225520208477974
1 0.0179920215159655
2 0.0146320248022676
3 0.0121880248188972
4 0.0104360226541758
5 0.00916802417486906
6 0.00837602093815804
7 0.00774402637034655
8 0.00725602637976408
9 0.00690001947805285
10 0.00659602647647262
11 0.00642402656376362
12 0.0063200332224369
13 0.00621201749891043
14 0.00612402195110917
15 0.0060280179604888
16 0.00600003264844418
17 0.00596802495419979
18 0.0059320330619812
19 0.00590002769604325
20 0.00589603185653687
};
\addplot [ thick, brown1363485, mark=diamond, mark size=1.5, mark options={solid,fill=none}]
table {%
-12 0.0990083590149879
-11 0.0906750038266182
-10 0.0861750245094299
-9 0.0757041946053505
-8 0.0677458569407463
-7 0.0622500330209732
-6 0.0555125288665295
-5 0.0464000217616558
-4 0.0385875217616558
-3 0.0334875248372555
-2 0.0267583578824997
-1 0.0220291912555695
0 0.0182125251740217
1 0.0148083521053195
2 0.0126708596944809
3 0.0112916901707649
4 0.0100043704733253
5 0.00880871620029211
6 0.00792611390352249
7 0.00732002267614007
8 0.00676403054967523
9 0.00632002111524343
10 0.00601202715188265
11 0.00581202516332269
12 0.005640032235533
13 0.0055160284973681
14 0.00538002979010344
15 0.00532402982935309
16 0.00513203395530581
17 0.00508802896365523
18 0.00503602484241128
19 0.00499603012576699
20 0.00497602950781584
};
\addplot [ thick, darkslateblue5134136, mark=o, mark size=1.5, mark options={solid,fill=none}]
table {%
-12 0.0968208536505699
-11 0.0886916890740395
-10 0.0848666951060295
-9 0.0772041976451874
-8 0.067637525498867
-7 0.061391681432724
-6 0.0545833446085453
-5 0.0464625246822834
-4 0.0404625274240971
-3 0.0330208614468575
-2 0.0261791944503784
-1 0.0212583560496569
0 0.0179250240325928
1 0.0153625188395381
2 0.0128333615139127
3 0.0110333533957601
4 0.00968263950198889
5 0.00840871967375278
6 0.00751306721940637
7 0.00681306235492229
8 0.00628802552819252
9 0.00588402757421136
10 0.00550402887165546
11 0.00526802055537701
12 0.00506002642214298
13 0.00490002613514662
14 0.0048080300912261
15 0.0046840263530612
16 0.00462002772837877
17 0.00453602569177747
18 0.00446402793750167
19 0.00440802332013845
20 0.00439603067934513
};
\addplot [ thick, darkorchid17068153, mark=triangle, mark size=1.5, mark options={solid,fill=none}]
table {%
-12 0.827332079410553
-11 0.81634795665741
-10 0.803140044212341
-9 0.78792405128479
-8 0.770119905471802
-7 0.748932063579559
-6 0.724088072776794
-5 0.695119917392731
-4 0.661803960800171
-3 0.623324036598206
-2 0.577759981155396
-1 0.52677994966507
0 0.4723239839077
1 0.414791941642761
2 0.356252014636993
3 0.298888027667999
4 0.245164036750793
5 0.198312014341354
6 0.159500002861023
7 0.129179999232292
8 0.107460044324398
9 0.0925720185041428
10 0.0821840167045593
11 0.074648030102253
12 0.0692240223288536
13 0.0654400214552879
14 0.0626360177993774
15 0.0607600212097168
16 0.059204027056694
17 0.0579720288515091
18 0.0569880232214928
19 0.0562480166554451
20 0.0557040274143219
};
\addplot [ thick, cadetblue68170153, mark=square, mark size=1.5, mark options={solid,fill=none}]
table {%
-12 0.825475990772247
-11 0.814328074455261
-10 0.801272034645081
-9 0.785607993602753
-8 0.767552018165588
-7 0.7464399933815
-6 0.721287906169891
-5 0.691788017749786
-4 0.657335996627808
-3 0.616692006587982
-2 0.571035981178284
-1 0.519244015216827
0 0.462575942277908
1 0.403191983699799
2 0.342276006937027
3 0.282188028097153
4 0.226688027381897
5 0.178548038005829
6 0.140664011240005
7 0.112216018140316
8 0.0923360288143158
9 0.0788400173187256
10 0.0691880211234093
11 0.062920019030571
12 0.0583880320191383
13 0.0551280304789543
14 0.05272002145648
15 0.0508200265467167
16 0.0494200214743614
17 0.0484080202877522
18 0.0476640313863754
19 0.0471120253205299
20 0.0467280261218548
};
\addplot [ thick, brown1363485, mark=diamond, mark size=1.5, mark options={solid,fill=none}]
table {%
-12 0.811033308506012
-11 0.798320829868317
-10 0.783958375453949
-9 0.764849901199341
-8 0.742558419704437
-7 0.717566728591919
-6 0.689029157161713
-5 0.65568333864212
-4 0.617604196071625
-3 0.574916660785675
-2 0.527800023555756
-1 0.476512521505356
0 0.42085000872612
1 0.364250034093857
2 0.307547867298126
3 0.252043485641479
4 0.203604385256767
5 0.162773936986923
6 0.130173921585083
7 0.107539147138596
8 0.0914956778287888
9 0.078248031437397
10 0.0687560141086578
11 0.0608760267496109
12 0.0548600181937218
13 0.0502920262515545
14 0.0468400195240974
15 0.0437800176441669
16 0.0414160303771496
17 0.0397160276770592
18 0.0383640229701996
19 0.0372680202126503
20 0.036472026258707
};
\addplot [ thick, darkslateblue5134136, mark=o, mark size=1.5, mark options={solid,fill=none}]
table {%
-12 0.808699905872345
-11 0.795529127120972
-10 0.780295848846436
-9 0.761954247951508
-8 0.740008294582367
-7 0.714120864868164
-6 0.684533357620239
-5 0.649233341217041
-4 0.609662532806396
-3 0.56484580039978
-2 0.514575064182281
-1 0.460312455892563
0 0.40326252579689
1 0.345687538385391
2 0.29023751616478
3 0.234356552362442
4 0.185226127505302
5 0.14506521821022
6 0.113900020718575
7 0.0917956829071045
8 0.0749960243701935
9 0.0635040178894997
10 0.0543400198221207
11 0.0477480217814445
12 0.0431120209395885
13 0.0397400222718716
14 0.0371240265667439
15 0.0352400280535221
16 0.0339880287647247
17 0.0327080190181732
18 0.0316600278019905
19 0.0307800248265266
20 0.0302040223032236
};
\end{axis}

\end{tikzpicture}}
        \caption{}
        \label{fig:over_SNR}
    \end{subfigure}
    \hfill 
    \begin{subfigure}{0.44\textwidth} 
        \centering
        \resizebox{\textwidth}{!}{
\begin{tikzpicture}
\pgfplotsset{%
    width=0.9\textwidth,
    height=0.95\textwidth 
}   
\definecolor{brown1363485}{RGB}{136,34,85}
\definecolor{cadetblue68170153}{RGB}{68,170,153}
\definecolor{darkgray176}{RGB}{176,176,176}
\definecolor{darkorchid17068153}{RGB}{170,68,153}
\definecolor{darkslateblue5134136}{RGB}{51,34,136}
\definecolor{lightgray204}{RGB}{204,204,204}

\begin{axis}[
legend cell align={left},
legend style={
  fill opacity=0.8,
  draw opacity=1,
  text opacity=1,
  at={(0.03,0.03)},
  anchor=south west,
  draw=lightgray204
},
log basis y={10},
minor xtick={},
minor ytick={0.0002,0.0003,0.0004,0.0005,0.0006,0.0007,0.0008,0.0009,0.002,0.003,0.004,0.005,0.006,0.007,0.008,0.009,0.02,0.03,0.04,0.05,0.06,0.07,0.08,0.09,0.2,0.3,0.4,0.5,0.6,0.7,0.8,0.9,2,3,4,5,6,7,8,9,20,30,40,50,60,70,80,90},
tick align=outside,
tick pos=left,
x grid style={darkgray176},
xlabel={SNR},
xmajorgrids,
xmin=-10, xmax=20,
xtick style={color=black},
xtick={-15,-10,-5,0,5,10,15,20,25},
y grid style={darkgray176},
ylabel={Task execution error rate},
ymajorgrids,
yminorgrids,
ymin=0.0036, ymax=1,
ymode=log,
ytick style={color=black},
ytick={0.0001,0.001,0.01,0.1,1,10},
yticklabels={
  \(\displaystyle {10^{-4}}\),
  \(\displaystyle {10^{-3}}\),
  \(\displaystyle {10^{-2}}\),
  \(\displaystyle {10^{-1}}\),
  \(\displaystyle {10^{0}}\),
  \(\displaystyle {10^{1}}\)
}
]
\node[draw=black, fill=white, rectangle, anchor=south] at (axis cs:11.5,0.29) {Task $2$};
\node[draw=black, fill=white, rectangle, anchor=south] at (axis cs:-4,0.017) {Task $1$};
\addplot [thick, darkorchid17068153, mark=triangle, mark size=1.5, mark options={solid,fill=none}]
table {%
-12 0.177356019616127
-11 0.163420021533966
-10 0.14859202504158
-9 0.134464025497437
-8 0.121536016464233
-7 0.108408033847809
-6 0.095116026699543
-5 0.0834600180387497
-4 0.0713920295238495
-3 0.0603360161185265
-2 0.0507840141654015
-1 0.0420640185475349
0 0.0349920243024826
1 0.0300080198794603
2 0.02533202432096
3 0.0221040267497301
4 0.0202120263129473
5 0.0182120259851217
6 0.0174400191754103
7 0.0160480216145515
8 0.0149760246276855
9 0.0146400211378932
10 0.0140760280191898
11 0.0136280301958323
12 0.0134480260312557
13 0.0132840275764465
14 0.0131320264190435
15 0.0127600310370326
16 0.0129480268806219
17 0.0131440237164497
18 0.0126600237563252
19 0.0127320261672139
20 0.012768030166626
};
\addplot [thick, cadetblue68170153, mark=square, mark size=1.5, mark options={solid,fill=none}]
table {%
-12 0.17796003818512
-11 0.164544031023979
-10 0.150640025734901
-9 0.137268006801605
-8 0.122980028390884
-7 0.10885201394558
-6 0.0951440334320068
-5 0.0823600292205811
-4 0.0700440183281898
-3 0.0589720159769058
-2 0.0489720292389393
-1 0.0407840237021446
0 0.0336920246481895
1 0.0282560251653194
2 0.0244960207492113
3 0.0206400267779827
4 0.0189520213752985
5 0.0166920181363821
6 0.0153320245444775
7 0.0147560313344002
8 0.0135680297389627
9 0.0132240224629641
10 0.0127440309152007
11 0.0124840261414647
12 0.0122120240703225
13 0.0119000319391489
14 0.0118720242753625
15 0.0116320280358195
16 0.0111760310828686
17 0.011368022300303
18 0.0115800313651562
19 0.0113200237974524
20 0.0115360235795379
};
\addplot [thick, brown1363485, mark=diamond, mark size=1.5, mark options={solid,fill=none}]
table {%
-12 0.0989480316638947
-11 0.094560019671917
-10 0.0932200327515602
-9 0.0907480269670486
-8 0.0810200273990631
-7 0.0747960209846497
-6 0.0680000111460686
-5 0.0600920282304287
-4 0.055024016648531
-3 0.045708030462265
-2 0.038800023496151
-1 0.0335240252315998
0 0.0305520202964544
1 0.0263560265302658
2 0.0237720273435116
3 0.0213400293141603
4 0.0192160326987505
5 0.0174760296940804
6 0.0157240275293589
7 0.0146680306643248
8 0.0135280275717378
9 0.0126480273902416
10 0.0122520234435797
11 0.0119280247017741
12 0.0109920259565115
13 0.0112320277839899
14 0.0107720233500004
15 0.0107600279152393
16 0.0103280283510685
17 0.0103680277243257
18 0.0100880265235901
19 0.0100560281425714
20 0.0100240232422948
};
\addplot [thick, darkslateblue5134136, mark=o, mark size=1.5, mark options={solid,fill=none}]
table {%
-12 0.0999440178275108
-11 0.0948680192232132
-10 0.0923240184783936
-9 0.08736402541399
-8 0.0800360292196274
-7 0.07505202293396
-6 0.065156027674675
-5 0.0557560175657272
-4 0.0498600192368031
-3 0.0447160191833973
-2 0.0381440259516239
-1 0.033516027033329
0 0.030104024335742
1 0.0252320226281881
2 0.0213800240308046
3 0.0179840177297592
4 0.0159640293568373
5 0.0147120309993625
6 0.0133840274065733
7 0.0123960236087441
8 0.0119400210678577
9 0.0110920239239931
10 0.0104240011423826
11 0.00980002153664827
12 0.00952402874827385
13 0.00899202562868595
14 0.00882400013506413
15 0.00880400184541941
16 0.00838400330394506
17 0.00821999832987785
18 0.00821600202471018
19 0.00805200077593327
20 0.00792400166392326
};
\addplot [thick, darkorchid17068153, mark=triangle, mark size=1.5, mark options={solid,fill=none}]
table {%
-12 0.834984064102173
-11 0.82563191652298
-10 0.814108073711395
-9 0.8021320104599
-8 0.787492036819458
-7 0.77018791437149
-6 0.749076068401337
-5 0.726500034332275
-4 0.698387980461121
-3 0.667315900325775
-2 0.631871938705444
-1 0.593383967876434
0 0.549316048622131
1 0.502984046936035
2 0.453935980796814
3 0.407716065645218
4 0.363347977399826
5 0.322924047708511
6 0.284951984882355
7 0.257420033216476
8 0.231867983937263
9 0.212412029504776
10 0.198755994439125
11 0.18511201441288
12 0.178412020206451
13 0.170400023460388
14 0.166220009326935
15 0.162780031561852
16 0.159984052181244
17 0.156548023223877
18 0.154304012656212
19 0.153000026941299
20 0.152232021093369
};
\addplot [thick, cadetblue68170153, mark=square, mark size=1.5, mark options={solid,fill=none}]
table {%
-12 0.830235958099365
-11 0.819640040397644
-10 0.808024048805237
-9 0.79338800907135
-8 0.777508020401001
-7 0.757407903671265
-6 0.734323978424072
-5 0.708504021167755
-4 0.677260041236877
-3 0.643180012702942
-2 0.603084027767181
-1 0.558632016181946
0 0.512188017368317
1 0.460488021373749
2 0.407788038253784
3 0.357376009225845
4 0.310200035572052
5 0.268084019422531
6 0.232292026281357
7 0.200156033039093
8 0.177964016795158
9 0.160352006554604
10 0.147280022501945
11 0.137228056788445
12 0.129292011260986
13 0.122820034623146
14 0.119888037443161
15 0.116244032979012
16 0.11419203132391
17 0.111220009624958
18 0.111032016575336
19 0.109728030860424
20 0.108480036258698
};
\addplot [thick, brown1363485, mark=diamond, mark size=1.5, mark options={solid,fill=none}]
table {%
-12 0.819132089614868
-11 0.808036029338837
-10 0.794407963752747
-9 0.780683994293213
-8 0.762987911701202
-7 0.742131948471069
-6 0.71974790096283
-5 0.692920088768005
-4 0.663880109786987
-3 0.630028009414673
-2 0.594396114349365
-1 0.555751979351044
0 0.514775991439819
1 0.469224005937576
2 0.422716021537781
3 0.379020005464554
4 0.335816025733948
5 0.299135982990265
6 0.264772027730942
7 0.237964034080505
8 0.213696002960205
9 0.191944003105164
10 0.178588002920151
11 0.164880022406578
12 0.156176000833511
13 0.147956028580666
14 0.139728039503098
15 0.133980020880699
16 0.128916025161743
17 0.12480802834034
18 0.123820021748543
19 0.119896031916142
20 0.116944029927254
};
\addplot [thick, darkslateblue5134136, mark=o, mark size=1.5, mark options={solid,fill=none}]
table {%
-12 0.813228011131287
-11 0.801060020923615
-10 0.785739958286285
-9 0.769420087337494
-8 0.749948024749756
-7 0.727539956569672
-6 0.702364027500153
-5 0.673672080039978
-4 0.643227994441986
-3 0.6078200340271
-2 0.567251980304718
-1 0.521408021450043
0 0.472792059183121
1 0.424280017614365
2 0.372891992330551
3 0.323756039142609
4 0.277400016784668
5 0.238900035619736
6 0.205512031912804
7 0.176212027668953
8 0.155460029840469
9 0.136920019984245
10 0.123116038739681
11 0.112007990479469
12 0.101927995681763
13 0.0951720029115677
14 0.0892879888415337
15 0.0841160044074059
16 0.0803839936852455
17 0.078247994184494
18 0.0758119896054268
19 0.0740879997611046
20 0.0728319957852364
};
\end{axis}

\end{tikzpicture}}
        \caption{}
        \label{fig:over_SNR_with_rotation}
    \end{subfigure}
    \caption{Task execution error rate over the SNR (a) without image rotation, and (b) with image rotation. 
    }
    \label{fig:over_SNR_both}
\end{figure}
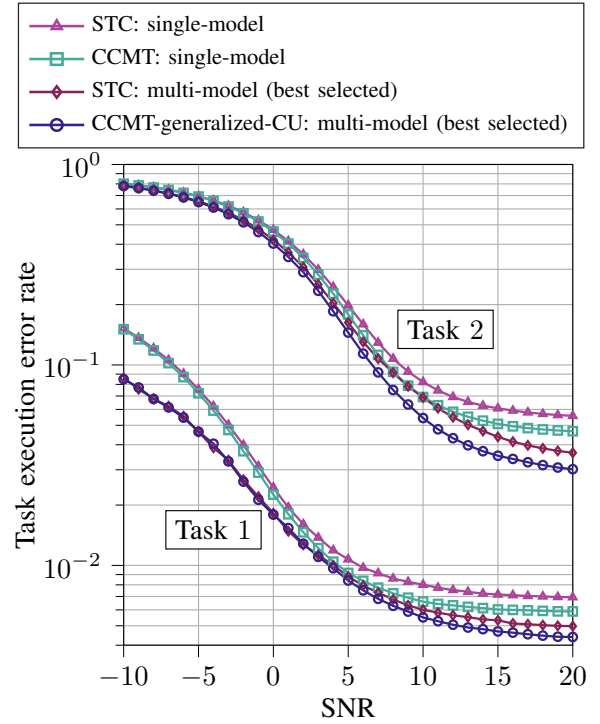
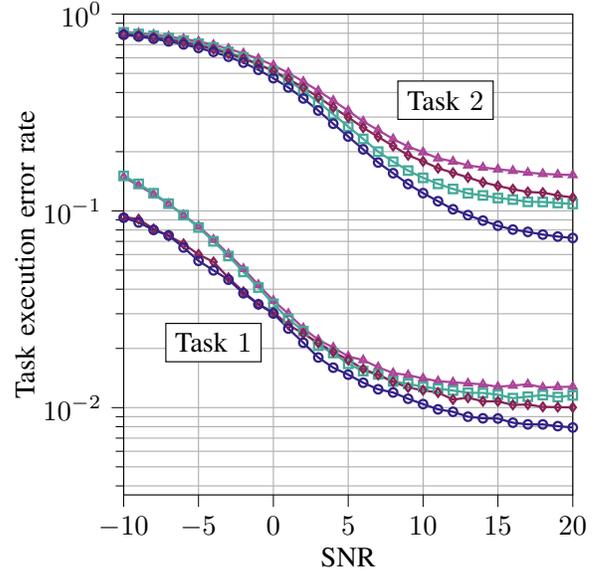

Further, the CCMT and STC are compared for different SNR values, in Fig. \ref{fig:over_SNR} without, and in Fig. \ref{fig:over_SNR_with_rotation} with image rotation, where each partial observation is individually rotated for an angle uniformly distributed between $\pm30^{\circ}$. For the simulation, multiple SUs models are trained for specific SNR ranges for the CCMT-generalized-CU and the STC, where in the evaluation process the best model is selected for each SNR. These are indicated by ``multi-model (best selected)" for both cases in Fig. \ref{fig:over_SNR_both}.
In total $11$ models are trained, with SNR ranges in $\text{dB}$: $[-12,-10],[-9,-7],$ $[-6,-4],$ $[-3,-1],$ $[0,2],$ $[3,5],$ $[6,8],[9,11],$ $[12,14],$ $[15,17],$ and $[18,20]$.

Moreover, the multi-model cases are compared with the ``single-model" cases, where only a single model is trained for the SNR range of $-10$ to $20 \, \text{dB}$ for the CCMT and the STC.

Fig. \ref{fig:over_SNR} shows that the multi-model cases (CCMT-generalized-CU and STC) outperform the singel-model cases (CCMT and STC) in the whole SNR range for task $1$ and for higher SNR values for task $2$. Next, the CCMT outperforms the STC for both the single and multi-model cases for higher SNR values. However, for lower SNR values, the CCMT and STC achieve almost the same task execution error rate, as the gain in performance from the CU in cooperative task processing is minimal compared to the errors introduced from the poor channel conditions.

Compared to Fig. \ref{fig:over_SNR}, Fig. \ref{fig:over_SNR_with_rotation} shows larger gaps in task error rate between the CCMT and STC architectures for both tasks in both single and multi-model cases. This indicates that the CU's cooperative processing of multiple tasks is more advantageous for more challenging datasets.

\subsection{Impact of Different NN Sizes}

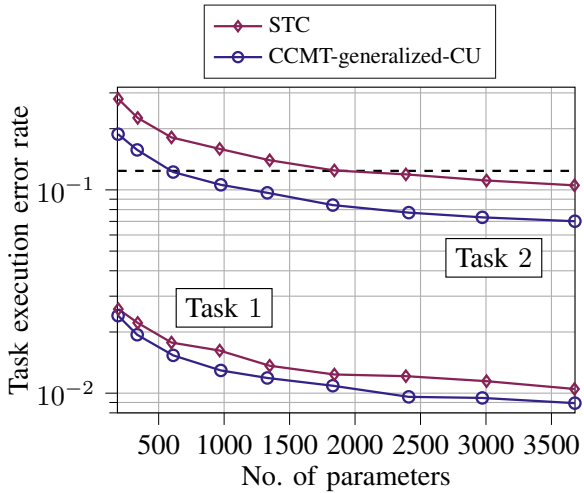
\begin{figure}[t!]
	\centering
    \vspace{-6.05cm} 
    \textcolor{white}{.}
    \resizebox{0.44\textwidth}{!}{\definecolor{brown1363485}{RGB}{136,34,85}
\definecolor{darkslateblue5134136}{RGB}{51,34,136}
\definecolor{indianred204102119}{RGB}{204,102,119}
\definecolor{skyblue136204238}{RGB}{136,204,238}

\begin{tikzpicture}
\begin{axis}[ %
 hide axis, 
xmin = 10, 
xmax = 50, 
ymin = 0, 
ymax = 0.4, 
legend columns = 1, 
legend style={at={(0.3,0)}, 
legend style={font=\footnotesize},
anchor = north west, 
draw = white!15!black, 
legend cell align = left}]

\addlegendimage{thick, color=brown1363485, mark=diamond, mark size=1.8, mark options={solid,fill=none}}
\addlegendentry{STC}

\addlegendimage{thick, color=darkslateblue5134136, mark=o, mark size=1.8, mark options={solid,fill=none}}
\addlegendentry{CCMT-generalized-CU}



\end{axis} 
\end{tikzpicture}}
    \vspace{0.1cm}
	\textcolor{white}{.}
    \resizebox{0.44\textwidth}{!}{
\begin{tikzpicture}
\pgfplotsset{%
    width=0.4\textwidth,
    height=0.31\textwidth 
}
\definecolor{brown1363485}{RGB}{136,34,85}
\definecolor{darkgray176}{RGB}{176,176,176}
\definecolor{darkslateblue5134136}{RGB}{51,34,136}
\definecolor{lightgray204}{RGB}{204,204,204}

\begin{axis}[
legend cell align={left},
legend style={
  fill opacity=0.8,
  draw opacity=1,
  text opacity=1,
  at={(0.91,0.5)},
  anchor=east,
  draw=lightgray204
},
log basis y={10},
minor xtick={},
minor ytick={0.0002,0.0003,0.0004,0.0005,0.0006,0.0007,0.0008,0.0009,0.002,0.003,0.004,0.005,0.006,0.007,0.008,0.009,0.02,0.03,0.04,0.05,0.06,0.07,0.08,0.09,0.2,0.3,0.4,0.5,0.6,0.7,0.8,0.9,2,3,4,5,6,7,8,9,20,30,40,50,60,70,80,90},
tick align=outside,
tick pos=left,
x grid style={darkgray176},
xlabel={No. of parameters},
xmajorgrids,
xmin=185, xmax=3680, 
xtick style={color=black},
xtick={500, 1000, 1500, 2000, 2500, 3000, 3500},  
xticklabels={ 500, 1000, 1500, 2000, 2500, 3000, 3500},  
y grid style={darkgray176},
ylabel={Task execution error rate},
ymajorgrids,
yminorgrids,
ymin=0.008, ymax=0.32,
ymode=log,
ytick style={color=black},
ytick={0.0001,0.001,0.01,0.1,1,10},
yticklabels={
  \(\displaystyle {10^{-4}}\),
  \(\displaystyle {10^{-3}}\),
  \(\displaystyle {10^{-2}}\),
  \(\displaystyle {10^{-1}}\),
  \(\displaystyle {10^{0}}\),
  \(\displaystyle {10^{1}}\)
}
]
\addplot[dashed, thick, domain=61:4000] {0.124};
\node[draw=none]at (0,0) {};
\node[draw=black, fill=white, rectangle, anchor=south] at (axis cs:3060,0.037) {Task $2$};
\node[draw=black, fill=white, rectangle, anchor=south] at (axis cs:1000,0.021) {Task $1$};

\addplot [thick, darkslateblue5134136, mark=o, mark size=2, mark options={solid,fill=none}]
table {%
190 0.0240640267729759
336 0.0193840265274048
611 0.0153440209105611
975 0.0129360174760222
1330 0.0118800234049559
1830 0.0108800223097205
2410 0.00959202088415623
2972 0.00947602279484272
3679 0.00893202051520348
};
\addplot [thick, darkslateblue5134136, mark=o, mark size=2, mark options={solid,fill=none}]
table {%
190 0.187960013747215
336 0.157240003347397
611 0.122368007898331
975 0.105900019407272
1330 0.096748024225235
1830 0.0841960161924362
2410 0.0772920250892639
2972 0.0732640251517296
3679 0.0701120123267174
};
\addplot [thick, brown1363485, mark=diamond, mark size=2, mark options={solid,fill=none}]
table {%
192 0.0259080212563276
340 0.0221720319241285
598 0.0177400261163712
966 0.0162040255963802
1348 0.0136428847908974
1842 0.0123571725562215
2388 0.0121240234002471
3004 0.0114560248330235
3676 0.010488023981452
};
\addplot [thick, brown1363485, mark=diamond, mark size=2, mark options={solid,fill=none}]
table {%
192 0.280043989419937
340 0.22598397731781
598 0.181172028183937
966 0.159072011709213
1348 0.139876022934914
1842 0.124620020389557
2388 0.119172021746635
3004 0.111224010586739
3676 0.10528402030468
};
\end{axis}

\end{tikzpicture}}
	\caption{Task execution error rate over the number of trainable CNN layer parameters. For task $1$ the SNR is $5\, \text{dB}$ and for task $2$ the SNR is $10\,\text{dB}$.}
    \label{fig:over_NN}
\end{figure}

Finally, the CCMT-generalized-CU and STC architectures are compared for different numbers of NN parameters. The task execution error rates are shown in Fig. \ref{fig:over_NN} over the number of parameters of the CNN layers of each sensing unit for the dataset with rotation. 

For this simulation. we consider different channel conditions for task $1$ and task $2$. For task $1$ the validation SNR is $5 \,\text{dB}$ with an SNR training range from $4$ to $6\,\text{dB}$ and for task $2$ the validation SNR is $10 \,\text{dB}$ with an SNR training range from $9$ to $11\,\text{dB}$. The number of convolution filters is increased from $c_1 = 4, c_2 = 2, c_3 = 2, $ $k_1 = 2, k_2 = 2, k_3 = 2$ resulting in $190$ and $192$ parameters, to $c_1 = 14, c_2 = 13, c_3 = 8, k_1 = 11, k_2 = 10, k_3 = 8$ resulting in $3679$ and $3676$ parameters, for the CCMT and the STC, respectively. The number of convolution filters in the final CNN layer $c_3$ and $k_3$ are always the same for the CCMT and STC. We note that CCMT and STC are trained for $500$ epochs in total for each case.

It can be seen that the CCMT saves a significant amount of computing resources compared to STC. For example, for task $2$, an error rate of about $0.12$ for the STC requires about $1900$ NN parameters, while the CCMT requires only about $611$. This is illustrated by the dashed line in Fig. \ref{fig:over_NN}. Moreover, we observe that in general, increasing the number of parameters decreases the error rate for all cases and the gap between the CCMT and STC stays relatively constant for the investigated parameter range. 

\section{Conclusion} \label{section.Conclusion}
We introduced the CCMT architecture based on an information-theoretic perspective, combining the cooperative processing of multiple tasks with the collaborative processing of distributed observations. We considered different training methods and channel conditions. Simulation results showed that the proposed CCMT architecture lowers the task execution error rate compared to the STC approach, specifically, for more challenging datasets and better channel conditions. Finally, it was shown that the advantage of the CCMT holds for different NN sizes.

\bibliographystyle{IEEEtran}
\bibliography{IEEEabrv,References.bib}

\end{document}